# Invisible Data Curation Practices: A Case Study from Facility Management

Tor Sporsem[1][0000-0002-5230-7480], Morten Hatling[1] and Marius Mikalsen[1][0000-0003-0882-7427]

[1] SINTEF, 7034 Trondheim, Norway
`lncs@springer.com`

**Abstract.** Facility management, which concerns the administration, operations, and maintenance of buildings, is a sector undergoing significant changes while becoming digitalized and data driven. In facility management sector, companies seek to extract value from data about their buildings. As a consequence, craftsmen, such as janitors, are becoming involved in data curation. Data curation refers to activities related to cleaning, assembling, setting up, and stewarding data to make them fit existing templates. Craftsmen in facility management, despite holding a pivotal role for successful data curation in the domain, are understudied and disregarded. To remedy this, our holistic case study investigates how janitors' data curation practices shape the data being produced in three facility management organizations. Our findings illustrate the unfortunate that janitors are treated more like a sensor than a human data curator. This treatment makes them less engaged in data curation, and hence do not engage in a much necessary correction of essential facility data. We apply the conceptual lens of invisible work – work that blends into the background and is taken for granted – to explain why this happens and how data comes to be. The findings also confirm the usefulness of a previously proposed analytical framework by using it to interpret data curation practices within facility management. The paper contributes to practitioners by proposing training and education in data curation.

## 1. Introduction

Buildings accounts for approximately a third of global energy end use and greenhouse gas emissions [1]. Norway has ambitions to achieve carbon neutrality by 2030, while the EU aims to decarbonize its buildings stock by 2050 and the building sector holds an important key to help reach these ambitious goals. The Norwegian sector of facility management addresses this challenge partly by pouring huge investments into digitalization with the purpose of operating and maintaining buildings smarter. Making informed decisions based on data of a portfolio of buildings is perceived as a way forward. Facility management companies look to gathering and utilizing data about their buildings to support smarter maintenance and more efficient operation. To succeed in establishing a smarter, more digitized process, the janitor plays an important part by holding responsibility of some of the data-input to the information system. Despite massive efforts, the building sector struggles to make its data valuable and ranks second at the bottom of Norwegian industries on digitalization according to McKinsey Global Institute Digitalization Index.

Naturally, responsibilities of front-line workers, such as janitors, profoundly change as additional attention is directed towards utilizing data. New tasks, like gathering and storing data about buildings they tend to, are required for realizing the sought digitalization effects. In addition to their traditional tasks, janitors become data curators. Data curation refers to activities related to cleaning, assembling, setting up, and stewarding data to make them fit existing templates [2]. Our focus in this study is how the curation practices of janitors – as front-line workers – fundamentally shape how data come to be and how their increasingly imposed role as data curators is performed [3].

As a step towards understanding how data curators' practices shape data and its use, we ask the following research question:

*RQ: How do invisible data curation practices shape how data come to be?*

We conducted a holistic case study [4] in three facility management organizations to examine how janitors record, curate and reuse data as part of a facility management system (FM system). Interviews and participatory observations were used to seek answers.

We draw upon an analytical framework of data curation practices [5] to interpret our findings. The characterization of data curation practices allows us to unveil how janitors shape data. Further, we conceptualize data curation practices as emergent and describe the role of curators. We seek to explain why data curation is challenging for front-line workers. One key reason for this is that new responsibilitiesza are imposed on front-line workers. Additionally, we show how the invisible work of janitors is crucial in data curation in facility management.

Janitors´ data curation practices are in effect invisible and taken for granted. This perspective on data curation as invisible work – which is work taken for granted and blending into the background [6] – explains the complexity of gaining valuable data and why workers such as janitors do not recognize data curation as "real work." A vicious cycle is created where those reusing data look elsewhere for high-quality information and those curating data downgrade its importance because of low use. Consequently, digitalization is viewed more like a burden than improvement.

Our study contributes to literature by providing empirical evidence on data curation in practice and offers insight into how invisible work impacts how data come to be. Additionally, it contributes by demonstrating how the analytical framework of Parmiggiani and Grisot [5] can support the facility management sector in unmasking mechanisms of how front-line workers record and reuse data. Theoretically, it provides one (of many more needed) confirmation of the framework's usefulness as a tool for both researchers and practitioners.

The rest of the paper is structured as follows: Section two introduces relevant literature on data curation and builds the conceptual lens of invisible work, followed by a case description and the research methodology in section three. Section four presents the main findings, of which the implications and importance are discussed in section five. We conclude with contributions to the body of knowledge and the practical problem of practitioners in section six.

## 2. Relevant theories and concepts

### 2.1 The need for understanding methods in data curation

Modern work practices, such as in the facility management sector, are becoming increasingly data centric. Buildings are automated, and decisions are made on data collected from the Internet of Things, amongst others [7, 8].

To understand data governance as data curation, we must analyze the ways in which data are curated. When engaging with the literature on data, Leonelli [9] finds that the novelty of data-centric approaches lies in (1) the prominence and status acquired by data as a scientific commodity and recognized output both within and beyond the field of science, and (2) the methods, infrastructures, technologies and skills developed to handle (format, disseminate, retrieve, model and interpret) data. These methods and infrastructures are essential because high-powered computations and new analytical techniques that automatically mine data, detect patterns and build predictive models have made it possible to deal with the abundance, exhaustiveness, variety, timeliness, dynamism, messiness and uncertainty of data [10].

### 2.2 Data work and data curation

The notion of knowledge infrastructures has been applied outside science to zoom in on the practical work of producing data. This is key, because the knowledge infrastructures in question not only mediate data but transform data in the process [12]. Thus, all pieces of the infrastructure matter and must be addressed in order for data governance to work. Studies of data work have shown how data are curated by continuously repairing cracks in the knowledge infrastructure [13], allowing the knowledge infrastructure to remain navigable (to search for data to use in analysis). Similarly, it is suggested that it is relevant to consider participation broadly in the study of data infrastructures [14].

To help us understand data work with a focus on how data curation practices shape data and data infrastructures, we look to the analytical framework of Parmiggiani and Grisot [5]. The framework specifies the unfolding of data curators' involvement and conceptualizes their work as emergent. Data curation practices contain three main activities in the framework: (1) *Achieving data quality* is described as practices for producing trustworthy data of sufficient quality for aggregated use. It concerns the fact that data quality depends on the skill and motivation of data curators, who can be well educated or hold no previous training, as well as on enriching data with additional data, for example by assessing the current data recording process with the one used for earlier data recordings. The activity of assessing data quality is shown to be situational and emerging and also connected with understanding the purpose of use of the data. (2)

*Filtering the relevant data* is described as practices for identifying data that can be useful for analysis both within and outside a context. Data filtering involves knowing which data are needed for what decisions and at what level, as well as knowing how to separate these data from the large amount of noise. As information infrastructure hold large amounts of data, this practice is crucial. (3) *Ensuring data protection* is practices that detect and flag possible threats to intellectual, technical and privacy property. The framework emerged from nine conceptual categories that surfaced during data analysis in Parmiggiani and Grisot's [5] case study.

### 2.3 Invisible work

How, then, can we understand these activities? A key aspect in understanding data curation is invisible work – because several curation practices are invisible. In [6], the authors summarize how work is invisible from different viewpoints and investigate how it influences the design and use of Computer Supported Collaborative work (CSCW) systems. They delineate three forms of invisible work: The first is creating a non-person, where the worker is treated more like a thing than a human. This is often associated with servants and domestic work. The second is disembedding background work as work that is expected as part of the background or infrastructure and that becomes invisible because of routine. This is often described through the example of nurses in hospitals. Such work is possible to observe if we look for it, but its nature of being taken for granted makes it invisible. The third form is abstracting and indicator manipulating, transforming work into measurements for productivity indicators. However, much work is invisible and impossible to measure, such as creative work.

Currently, the facility management sector is not equipped with tools or frameworks to understand the emerging curation practices. We apply this framework as a conceptual lens to make sense of our data and simultaneously demonstrate how this framework can contribute to a sector trying to comprehend the extent of its digital transformation.

## 3. Method

### 3.1 Case description

Facility management consists of administering, operating and maintaining buildings that we all use, such as office buildings, residences, stores, schools, etc. Large companies within facility management in Norway are exploring new ways of retrieving and using data from buildings. The rational is to save money and climate footprint through data-based decision-making in operating and maintaining buildings.

This case comprises a private facility management company, the municipality of Molde, and Møre og Romsdal County. These organizations established a joint project to investigate the potential of unexploited data that already exists in their standing buildings. All buildings are situated in the city of Molde, Norway. Additionally, two local technology start-ups participated; their rationale was to contribute technology competence while simultaneously gaining insight into the facility management sector.

The authors of this paper were included in early ideation and participated as research partners throughout the project period of 06.2020-01.2021. Research funding was provided by the Norwegian state through its company, Innovation Norway.

The three facility management organizations are similar in their way of operation but hold different sizes of building portfolios. All three hold office buildings and residential houses, while the municipality exclusively holds elementary schools, institutions for the elderly and kindergartens, and the county municipality holds high-schools. They use off-the-shelf FM systems to manage with the purpose of integrating all data about their buildings to gain insight through data analysis. Both static (i.e. documentation, technical drawings, warranties, etc.) and dynamic data (i.e. electric current usage, maintenance history, tenancy contracts, etc.) are supposed to be included and updated live. The intention is to enable decision makers to make well-informed decisions based on actual data and offer a tool to support maintenance planners and workers (such as janitors). All three organizations use different off-the-shelf FM systems modified to fit their individual needs. Implementation of the FM systems took place only 1-2 years ago for all three organizations, and is ongoing, as they are still adjusting and finding ways of using them. We view these FM systems as data infrastructures, similar to the definition of [15].

Janitors hold practically the same responsibilities in all three organizations. They are the building's caretaker, constantly monitoring its condition, conducting small maintenance work, supporting tenants in facility issues, reporting issues that require professional support, coordinating third-party craftsmen, and curating data for managers' decision-making processes. A janitor's schedule is usually flexible in order to meet unexpected daily maintenance, tenants' changing needs, inspection deadlines from checklists and external happenings such as snowfall. The biggest difference in tasks between the organizations is the level of dependence on their FM system. The municipality and county are slightly more advanced in their use, as they have integrated maintenance planning. As information infrastructures, like the FM systems, are implemented, janitors are effectively becoming data curators because they are expected to conduct curation tasks to serve the FM systems as part of their work activities. Naturally, new practical problems emerge in the wake of the new data curation tasks.

### 3.2 Research design, data collection and analysis

As janitors' line of work very similar in all organizations, we decided to use a holistic case study method, as described by Yin [4]. We maintained an exploratory approach, as we did not set out to test any specific theory or hypothesis [16]. The unit of analysis was the three organizations in the facility management sector. More specific, janitors conducting data curation and managers as data users. We hold an interpretive view in this study, comprehend the world and its truths as subjective realities [17].

Data collection was performed in rounds, spanning three months, conducted by two researchers in collaboration, summarized in Table 1. We started off with an exploratory mindset to smoke out practical problems. The first workshop offered insight into the partners' concerns and helped form a semi-structured interview guide. Then three interviews were conducted with the purpose of further understand what problems were relevant researching, and yielded 11 transcribed machine-written pages. This way of iteratively considering the independent meaning of parts and the whole that they form – called the Hermeneutic Circle [18] – led us towards the research question and enabled us to create observation guides.

Participatory observations were conducted by both the first and second author, tracing four janitors for four hours each during a random workday and joining them for lunch. Observations spanned two days, during which the researchers split up, following one janitor each in parallel and meeting up for reflections and discussions afterwards. Our impression is that the janitors appreciated showing their work practices and expressed themselves freely. We felt more like apprentices than researchers (we also wore similar clothing as the janitors to lessen the researcher-subject gap). As our research material was socially constructed through interaction between researcher and subjects, we questioned each other about our assumptions to trigger reflections [18]. Research notes and pictures were taken during observation, and reflections were written immediately afterwards, resulting in 8 pages of machine-written notes produced.

The first and second author jointly analyzed gathered data and immersed themselves in the material. A word processor software was used for both open ended coding and memoing [19]. An example of a code label is "training and support", with codes "support functions to janitors are not present in the FM system" and "all experience/knowledge is based on operational trial and error." Our labels were then compared to the category in Parmiggiani and Grisot's analytical framework [5], which functioned as a critical look on our analysis. Our findings were then presented and discussed in a workshop involving janitors and facility managers in all three organizations to check if our findings represented the world as they know it and adjust any misapprehensions.

**Table 1.** Data Sources

| Data source | Description | N |
|---|---|---|
| Semi-structured interviews | We interviewed one manager from each organization who is responsible for the work of janitors and supports them in maintenance planning and budgeting | 3 |
| Observations | We participated in the normal day of four janitors, two from the private organization and one each from the municipality and county. Each janitor was observed for four hours. | 16 hours |
| Workshops | First, we held a physical kick-off meeting with all partners participating (12 participants). The problem statement and partners' perspectives and expectations were discussed. In the second workshop, we validated preliminary findings through discussion and obtained feedback supporting our analysis (17 participants) | 2 |
| Documents | For all three case organizations, organizational charts, janitors' work procedures and checklists were examined. | 27 pages |

# 4. Results

In this section, we present our findings on janitors' data curation practices and their use of the FM system. We end the section by showing how our findings relate to the analytical framework's categories in Table 2.

**4.1 Janitors do not register valuable data in the FM system**

The three interviewed property managers highlighted the importance of up-to-date data as crucial for predictability in maintenance budgets and property investments, and they acknowledged that janitors hold a key role in recording these data. Hence, the purpose of the FM system is to maintain an updated decision basis for planning, maintenance and investments. At the same time, the FM system is also supposed to be a support tool for operational personnel such as janitors to help them find all relevant building information for the task they are doing. One property manager explained his ambitions for the system's relevance to janitors, "Our goal is for janitors to find everything they need to know about a building in the FM system […] like historical maintenance, earlier observations, technical documentation, you name it. […] I extract data and send it to those planning maintenance and property investment." The janitors, on the other hand, did not ascribe the same level of importance of data in their daily work and did not find the FM system as supportive as intended by property managers. Janitors generally expressed a lack of understanding of the purpose of registering data and consequently the relevance to their work: "I don't need an updated FM system. I've got full control [of information] of my buildings."

Typically, janitors did not record or correct data related to unplanned tasks or maintenance, such as replacing a fluorescent lightbulb or tightening a leaking water tap. Both janitors and property managers regard such tasks and maintenance as crucial in keeping their tenants satisfied, but at the same time, they also acknowledge that as long as everything is working, few are interested in what they do. Despite being recognized as the most important type of work, the janitors do not find value in recording it. The data has little practical use in their day-to-day operations, and they fail to see how this data is valuable to others: "My reporting that this light bulb has been replaced has no value. No one uses that information. It only requires extra work [in recording the data]." As long as this data is unavailable to property managers, they are unable to use it in maintenance planning. When asked, property managers describe such data as "incredibly useful data".

As a response, janitors were offered more user-friendly technology for recording and correcting data. For example, they were encouraged to take pictures of completed work and upload them via the FM app on their phone, without any other recording. This did not lead to any improvement. A janitor said, "It is quite easy to just take a picture, but I forget to do it as it has no importance." This suggests that simplifying data recording technology will not raise the level of data recordings (it is difficult to think of any easier way of recording than taking a picture). Rather, understanding the purpose and importance of data for the organization as a whole motivated and seemed to increase the janitors' amount of recording data into the FM system. Additionally, the quality of the recorded data increased when the janitors knew the reuse purposes of data. It appears that neither the FM system nor the property managers focused on showing the reuse purposes to janitors.

Our study also found that janitors prioritized recording data they themselves found useful. For instance, they recorded data on how they performed complicated maintenance, and observations on critical hardware in their buildings. Data that, in some cases, were outside the scope of the current digitized FM system; in other cases, the janitors chose to not use the FM system. However, we found several examples of how the janitors recorded such data in notebooks, Post-it notes, Excel files, ring binders, etc. – outside the FM system. When asked why they preferred to do it like that, janitors explained that recording and navigating to retrieve data is time-consuming when using the FM system. Further they claimed their ways of storing data held higher usability. For example, we followed a janitor who built and maintained a separate Excel document with a list of all the filter types used in all of the ventilation systems within his building portfolio, even though the same information was stored in the FM system. The reason he gave us was that the FM system required the janitor to visit each individual object to retrieve filter data to the annual ordering of new filters, which would require hours of work. "Instead, I send my entire [Excel] list to the supplier and have thus placed the order in just a few minutes." He frequently updated this list with new filter data instead of updating the FM system. This showed low local usability kept data hidden and unavailable for others to reuse. Another janitor we observed kept old paper-based procedures for maintaining a firefighting system because it featured a comment field where he recorded data about unusual occurrences. He showed us that the same field did not exist in the checklist in the FM system. Thus, the data only existed outside the FM system – a system which failed to acknowledge this important practice of noting comments.

These findings show that janitors' data curation practices in effect keep data out of the FM system and consequently out of reach for others to reuse. Further, it shows how the usability of the FM system drives janitors to filter and avoid correcting data, hence shaping their data curation practices. As a result of this, we found that property managers contact janitors to gain access to these data for decision-making processes (if they know the information exists at all).

### 4.2 The FM system does not add value to janitors

None of the janitors we followed felt the FM system offered information or knowledge they did not already have. By being present in their buildings, they catch all the information they need by using their senses and listening to rumors. They capture vast amounts of data through listening, smelling, touching and looking. For example, a strange smell in a technical room may indicate overheating of equipment, and a discoloration may be the accumulation of condensation. They make such observations all the time in enormous numbers; they are constantly updated on the condition of the buildings.

In addition, janitors pick up rumors through presence. For example, we observed a janitor who, through conversation, picked up information about moving activity of one of the tenants. The janitor had a suspicion that the moving activity would generate clutter on the pavement outside the premises, which turned out to be true. "I do not need an updated FM system, I have control of my buildings anyway," he says. Janitors' updated and detailed status of their buildings' conditions are far from matched by the FM system and thus the FM system does not offer them any added value or incentive to use it.

**4.3 Janitors hide their work to avoid further misfitting of the FM system**

A substantial part of janitors' tasks consists of proactively inspecting buildings and installations to plan maintenance, correct errors early and avoid breakdowns. This work is traditionally organized through checklists that describe what janitors need to inspect and when. One janitor we followed explained that he is always searching for opportunities to do inspections while conducting other work throughout his days. For example, after fixing a toilet in an office building, he inspects a nearby water-heater room and emergency lighting in a stairway before leaving. In this way, he uses the available flexibility within his schedule to do most inspections required by checklists when he visited buildings anyway, saving extra travel time.

However, the janitor expressed concerns over checklists being digitalized and part of the FM system. The purpose, he thought, was to obtain better data quality. But it had resulted in reduced flexibility for one of his colleagues. Tags had been put up to be scanned when an inspection was done to provide a timestamp. This resulted in his colleague having to drive back and forth to visit each building to sign off inspections he had already done while conducting other work. "This is a way of making up work for ourselves," he said. The janitors were concerned that the FM system was a gate-opener for similar initiatives, leading them to hide inspection work from their managers to avoid drawing unnecessary attention to the checklists. Janitors fear the FM system can be further misfitted to their local needs and try to steer their managers' attention elsewhere to avoid any "suggestions for improvement."

**4.4 Categorizing the findings**

To make sense of our findings, we analyzed them through the nine constructs in Parmiggiani and Grisot's analytical framework [5] and sorted them into the three categories of data curation practices, as shown in Table 2. Because we did not connect any of our findings to the category of *ensuring data protection*, it is empty in the table and not given further consideration in this paper.

**Table 2.** Findings in relation to Parmiggiani and Grisot's analytical framework categories

| Concepts | Findings |
|---|---|
| Achieving data quality | Janitors did not know the purpose of reuse and therefore could not assess its usability and quality for property managers |
|  | Janitors were not incentivized to record more data (enrich) than decreed, because the FM system offer little added value to janitors locally in terms of knowledge or information |
|  | Janitors did not correct errors they observed in data sets because they did not perceive the data as valuable or purposeful |
| Filtering the relevant data | Janitors did not perceive data form some of their tasks (i.e., changing lightbulbs) as important enough to record it in the FM system |
|  | Janitors recorded data in their own "home-made" systems instead of the FM system because the FM system had poor usability and was time consuming to use. |
| Ensuring data protection | No findings |

## 5. Discussion

In this chapter, we answer our research question by discussing how janitors are perceived more similar to how sensors produce data. We also discuss how this treatment influences their data curation practices. As a reminder to our reader, we return to the research question: *How do invisible data curation practices shape how data come to be?* Finally, we offer the lens of invisible work to explain why data curation practices are fundamentally challenging when producing data.

In our case, Janitors are not recognized as data curators but treated as a homogenous group of mere data providers – the same way as a sensor automatically delivers raw, untampered data. This leads to several misconceptions. There is no need to explain the use of data to a sensor, provide feedback on its practices or provide training and education. Our findings show however, that janitors do several curation activities such as filtering and enriching data. If janitors are not recognized as data curators, they are effectively turned into non-persons [6]. In the following paragraphs, we elaborate on this point.

The purpose of reusing data is not explained to janitors by either managers or the FM system. Thus, janitors filter out data because they do not realize its usefulness and they save time by not recording. As a consequence, essential data on day-to-day maintenance are seldom enriched or even recorded, and the information never reach data users, which are property managers in this case. One case organization tried to solve this problem by introducing easier-to-use technology. However, reuse purposes of data were not conveyed, and janitors remained demotivated to record data. The failure of this attempt shows us that one rather should recognize janitors as humans who need to perceive their work as meaningful, not only convenient.

Training and education in data curation is not provided to janitors if they are treated like sensors. Earlier research has shown that curators constantly need to make time to learn new skills to curate data [5, 20]. Our findings agree that this should be met with increased training where curators can learn the craft of curating data and the purpose of reuse. Additionally, curators such as janitors should receive training on ways to leverage data to meet their own data needs. This becomes especially important when we consider that janitors usually do not have any formal education or skills in data curation.

We suggest acknowledging that janitors' data curation is invisible work [6]. As we have seen, janitors are treated like a sensor instead of a human data curator. This fits the description of invisible work in which people are treated more like things than as humans, creating nonpersons. "Under some conditions, the act of working or the product of work is visible to both employer and employee, but the employee is invisible" [6] p. 15. Of course, janitors are not wholly treated like nonpersons by their organizations, and they perform many recognized tasks. However, our findings show data curation is not one of them.

Further, janitors' curation practices fit a second form of invisible work, which is expected as part of the background or infrastructure [6]. If one went *looking* for it, then it is physically possible to observe it, but since it is taken for granted by others, it is functionally invisible. In our results, janitors' curation work is comparable to findings on nurses who make data choices when recording patient journals, effectively doing curation work of filtering and enriching (or impoverishing) [5]. Nursing is a commonly used example of invisible work happening in the background [21].

We argue that recognizing this work as invisible is the first step towards making it visible and bringing it to the attention of organizations and their information systems. We agree that foregrounding data curation practices can lead to necessary involvement in reuse and provide training [5]. As long as janitors are unrecognized as curators by management, data users or data infrastructures, their stance will not improve, and they will not be fully able to do their job as data curators.

Our findings suggest that *data curation* is a fitting concept explaining what janitors do with data, and that comprehending its implications can be fruitful for facility management. We have shown that janitors' data curation practices are invisible. In continuation, we discuss how janitors' curation practices affects the ways data come to be and how it fundamentally shapes data.

Data are treated as raw and complete when data curation is invisible. Plantin [22] and Parmiggiani and Grisot [5] argue that data curation practices, which are invisible to managerial levels, should be accounted for to erase the misconception of data as raw in decision-making processes. Our findings support this in showing that curators make decisions about data on a day-to-day basis – often under the radar – which fundamentally contributes to shaping data [5].

This serves as an example of Jones' [23] point about understanding the practices that are involved in creating data: "Data is partial and contingent and brought into being through situated practices of conceptualization, recording and use,". Data then, are not simply referential, natural and objective representations of the world. *How data come to be* (how it is produced) contains three steps as follows: (i) what data about the real world *can* be recorded, (ii) what gets chosen to be recorded, and (iii) what actually gets recorded [23]. We argue that invisible data curation practices are one important factor of these *situated practises* in that they are regarded as an implicit factor – so high degree of implicitness that it becomes invisible. This is one explaining factor to why we are not able to obtain, raw, untampered, objective data about the world. In reuse (e.g., making decisions, maintenance planning and budgeting), correct data are crucial to success. When data do not represent the real world, decisions, plans and budgets will not either.

Thus far, this discussion has shown that janitors' data curation practices can be understood as invisible work and that this fundamentally shapes data and data curation practices. To end this section, we argue that janitors will not include data curation as part of their primary work if curation work continues to be treated as invisible.

One of the case janitors described their responsibility in data curation work as "not real janitor work." He argued that the FM system did not offer value in doing their primary tasks. Neither managers nor the FM system acknowledged their local needs for data to support their daily work. As we have shown, this results in data infrastructures that are not tailored to janitors' reuse of data. In effect, the infrastructure only recognizes the global data needs of managers and undermines the local data needs of the curators. However, our findings show that janitors are willing to do data curation work if they see it benefit themselves.

Hidden in the shadows, janitors make local systems that fit their needs, one size fits one, such as personal notebooks and Excel-files that they tailor to their own needs. This seemed to be well known by managers, who sought these local systems to obtain important information to support their decision-making. The local systems end up attracting most of the curators' attention, and they are frequently updated with high-quality data. A vicious cycle occurs in which janitors do not update the FM system with relevant data and reuse gives little value, which leads to data users finding other ways to obtain needed data, thus leading to even lower use.

The issue raises the question of whether janitors have ever been asked their opinion about including data curation work as part of their job description, or whether this is something imposed on them. Having to spend more time pleasing global data needs and reuse reduces their time for "real janitor work." We have already shown that janitors wanted to keep their inspection checklists hidden to avoid having to record them in the FM system. This is like the way nurses have struggled to keep their work ambiguous and discreet to avoid cumbersome paperwork [24]. Wagner [24] showed that nurses feared more visibility may lead to more surveillance. Recording data represents a potentially new way of surveillance and control, whereas in our case, janitors pointed to the task of having to register tags when conducting inspection work as a form of surveillance and control. As such, exposing data curation practices represented a threat to their autonomy rather than a tool to improve their day-to-day work.

After interviewing and observing janitors, we suggest that recognizing invisible data curation practices will support janitors in understand reuse purposes, reveal janitors' local reuse needs, and show data users that data are not raw and complete. This can unlock their potential to become high-performing data curators.

## 6. Conclusion

In this paper, we examined how janitors' data curation practices fundamentally shape data in the facility management sector. Through the holistic case study of janitors as data curators in three organizations, we showed that they are treated similar to sensors instead of curators. Additionally, we showed that this treatment influenced their curation practices to filter out, rather than correcting or enriching essential data. We also showed that they perceive curation activities as a burden, not as "real janitor work."

This paper offers an explanation of these phenomenon in the facility management sector by applying the lens of invisible work [6], as suggested in earlier research [5]. We found this to be a comprehensive way of understanding the world of a janitor that enabled us to offer a novel understanding to our case organizations. This understanding was well received. "It is not nice reading for a manager; however, this is incredibly valuable feedback," said the facility manager of the Møre og Romsdal County case organization.

Our study contributes to the theoretical field of data curation by demonstrating that Parmiggiani and Grisot's analytical framework [5] useful to understand data curation practices and to reveal invisible curation practices. Future research is needed to explore the applicability of the framework in other contexts and sectors. Further, our study propose implications for the discourse on datafication [23] and the understanding of how situated practices shape data and its reuse. It also supports earlier suggestions to understand data curation as invisible work [5, 22].

The study has implications for practitioners such as janitors. First, by unveiling invisible curation practices we show the need for training and education of front-line workers in data curation. Today, they are expected to handle data curation work without offering them support or guidelines. Educated and motivated curators will contribute to higher value data. Second, by describing invisible data curation as a fundamental challenge in producing data so they can confront the correct problem. We urge designers of data infrastructure such as the FM system to recognize curators' local data needs and their invisible curation practices. In the case of facility management, a point of departure would be to analyze the features janitors' build into their "homemade" local systems because they both represent their needs and reveal their invisible curation practices.